# Morphology of Graphene on SiC($000\bar{1}$) Surfaces

Luxmi, P. J. Fisher, N. Srivastava, and R. M. Feenstra
Dept. Physics, Carnegie Mellon University, Pittsburgh, PA 15213

Yugang Sun
Center for Nanoscale Materials, Argonne National Laboratory, Argonne, IL 60439

J. Kedzierski and P. Healey
MIT Lincoln Laboratory, Lexington MA 02420

Gong Gu
Sarnoff Corporation, CN5300, Princeton, NJ 08543

**Abstract**
Graphene is formed on SiC($000\bar{1}$) surfaces (the so-called *C-face* of the crystal) by annealing in vacuum, with the resulting films characterized by atomic force microscopy, Auger electron spectroscopy, scanning Auger microscopy and Raman spectroscopy. Morphology of these films is compared with the graphene films grown on SiC(0001) surfaces (the *Si-face*). Graphene forms a terraced morphology on the C-face, whereas it forms with a flatter morphology on the Si-face. It is argued that this difference occurs because of differing interface structures in the two cases. For certain SiC wafers, nanocrystalline graphite is found to form on top of the graphene.

Graphene, a monolayer of carbon atoms, has received tremendous attention in the last few years because of its exceptional electronic properties.[1] High electron and hole mobilities, long coherence length and ballistic transport in this material makes it a promising candidate for future device applications.[2] There are different ways of obtaining graphene, such as micromechanical exfoliation of highly ordered pyrolytic graphite (HOPG) onto a solid support,[1] segregation or deposition of carbon from/to metal surface[3] or thermal decomposition of SiC crystal under vacuum conditions.[4,5,6] Due to the availability of large-area semi-insulating SiC substrates, the latter method is advantageous for device and circuit applications of graphene.[7]

There are several differences in the graphene grown on the two polar faces of the SiC i.e. the (0001) *Si-face* or the ($000\bar{1}$) *C-face*. It is easier to graphitize C-face samples,[5] thicker films are formed, and also there is rotational disorder in graphene films grown on the C-face.[4,6,8] Some research groups have shown that the method of annealing in high or ultra-high vacuum does not produce homogeneous graphene on the C-face.[7,9] In this paper, we demonstrate that annealing in high vacuum can indeed form high quality graphene on the C-face. Field effect mobilities of the transistors made from these samples exceed 4000 $cm^2$/Vs at room temperature.[10] We do, however, find that the morphology of these films, as seen in atomic force microscope (AFM), is quite different from the Si-face graphene films. While the graphene films have a relatively flat morphology (with small pits) on the Si-face, the films form with a terraced morphology on the C-face. The graphene is found to be *thinner* on the upper terraces as compared to on the lower terraces. We also find that, for certain wafers, a significant coverage of nanocrystalline graphite (NCG) forms on top of the graphene films.

Samples measuring 1×1 $cm^2$ were cut from 2 or 3 inch diameter "epi-ready" SiC wafers obtained from Cree Inc., using both semi-insulating 4H-SiC and n-type 6H-SiC C-face material. Samples are ultrasonically cleaned in acetone and methanol before introducing them into the vacuum chamber. Most samples were annealed using a resistively-heated graphite strip, as previously described,[11] although in a few cases direct resistive heating of the n-type SiC was employed, with identical results. In order to remove polishing damage on the as-received wafer surfaces, samples are annealed in a 1 atm hydrogen environment at 1500°C for 3 min.[12] This procedure produces an atomically flat surface, with uniformly spaced steps arising from unintentional miscut of the wafer. Following this procedure the hydrogen is pumped away and the samples are annealed in vacuum at temperatures above 1100°C, for 20 min, with a background pressure of $10^{-8}$ Torr (mainly hydrogen). The thickness of graphene films is determined by Auger electron spectroscopy[11] using the C KLL line at 272 eV and either the Si KLL line at 1619 eV or the Si LMM line at 92 eV; the relatively long escape depth of the former allows determination of thicknesses up to ≈20 monolayers (ML). Calibration of the Auger sensitivity factors is accomplished using graphene thicknesses determined by low-energy electron microscopy (to be discussed elsewhere), with accuracy of about ±20%. Low-energy electron diffraction reveals rotated stacking of the graphene layers similar to that previously seen,[13] although the rotation angles in our case are 30°±φ with φ ranging from 6° to 13° as opposed to the value of φ=2.2° previously reported.[13]

Figure 1 shows AFM images of samples annealed at different temperatures. After annealing at about 1120°C, we can see in Figs. 1(a) and (b) that the overall step-terrace pattern, as seen after H-etching,[12] is maintained along with some small changes in the morphology. On the terraces, small domains (≈200 nm in extent) of varying gray-contrast are seen; these domains are similar to those reported in scanning tunneling microscopy studies of C-face graphene, and attributed to differing underlying interface structure between the graphene and the SiC.[14] After annealing at 1190°C, the morphology further evolves, forming raised and lowered terraces as seen in Fig. 1(c). The small domains are still faintly seen, and the shape of the terraces now deviates from the original step structure of the H-etched surface. It should be noted that those original steps can still be discerned in Figs. 1(a)-(c) by the lines of white deposits [≈1 nm high, and vertically aligned in Fig. 1(c)]. After annealing at 1240°C, Fig. 1(d), the surface is fully transformed into a morphology with distinct terraces of differing heights. All the terraces still have faint white lines extending over them. These lines perhaps delineate domains in the graphene, or alternatively they could be related to the larger (higher) raised ridges or "puckers" seen in Fig. 1(e). In that image, obtained after annealing at 1320°C, the puckers are believed to arise from thermal expansion mismatch between the substrate and the graphene films.[15]

The features seen in Fig. 1 for graphene on C-face SiC films are quite different than those observed on Si-face graphene films.[11,15,16,17] There, small pits are seen on the surface, even during the early stages of graphitization. The origin of these pits is recently explained in terms of the development of the $6\sqrt{3} \times 6\sqrt{3}$-R30° (which we denote as $6\sqrt{3}$, for short) interface layer between the SiC and the graphene.[17] This interface layer apparently acts as a template for subsequent graphene formation, with the graphene forming from the $6\sqrt{3}$ layer and the $6\sqrt{3}$ structure moving *downwards* as Si sublimates from below it.[8] For the C-face, however, the graphene/SiC interface can consist of various different structures,[14] and apparently any sort of uniform templating behavior analogous to that of the $6\sqrt{3}$ does not occur. On the contrary, we hypothesize that the different graphene/SiC interface structures observed to occur for the C-face[14] may have varying efficacy for producing graphene. The interface structure for the C-face would thus still act as a template for Si release and resulting graphene formation, but different structures would have varying rates for this process. Different thicknesses of graphene would form on the surface, at least initially, which could account for the varying contrast seen in the small domains of Figs. 1(a)-(c).

To further probe the nature of the terraced morphology seen for the C-face, scanning Auger microscopy studies were carried out. Figure 2(a) shows an AFM image of a C-face graphene film and Fig. 2(b) shows a scanning electron microscope (SEM) image from the same sample. The morphology as seen by AFM consists of raised terraces over a minority of the surface area, similar to that of Fig. 1(d). Examining the SEM image, we see that it displays just the same type of morphology, and we therefore associated the white portions of the SEM image with raised terraces. Now, examining the individual spectra of Fig. 2(c), acquired from the locations marked in Fig. 2(b), we see that the white (raised) area in the SEM image shows a Si peak with greater amplitude than that from the darker (lower) area on the surface. Similar results were obtained at multiple locations over the surface, and we conclude that the *raised* areas in the topography actually have a *thinner* coverage of graphene, with the *lower* areas having *thicker* coverage. This conclusion is consistent with an expectation of a *lower* morphology and concomitantly *thicker* graphene in areas where more Si has left the surface, i.e. assuming that the Si leaves the SiC by diffusing through the overlying graphene without significantly redistributing itself. This result is also consistent with the above hypothesis of having different interface structures between the graphene and the SiC which produce differing graphitization rates for different areas of the surface.

We mention here one additional aspect of the morphology of the C-face graphene films that we have observed for certain wafers. Figure 3(a) shows a sample prepared in an identical manner as described above. This sample displays two types of surface morphology: a low (black) topography covered by small domains, similar to those of Fig. 1(d), and a high (light gray) topography. The latter is quite disordered and rough, as seen in the inset of Fig. 3(a). The optical micrograph of Fig. 3(b) is taken from the same sample as Fig. 3(a), and the characteristic hexagonal-shaped disordered areas are again evident. Figure 3(c) shows spatially resolved Raman spectra, one acquired from within a disordered area inside a hexagon and the other from an ordered area between hexagons. Both spectra show characteristic peaks associated with both SiC and graphene,[11,18] the latter marked as D, G, and 2D. However, the spectrum from the disordered area differs from that of the ordered area in that it reveals a higher intensity of the defect-related D peak at 1360 cm$^{-1}$ and its G peak is shifted by $\approx$10 cm$^{-1}$ to 1596 cm$^{-1}$. This type of shift is known to be associated with the occurrence of nanocrystalline graphite (NCG),[19] and we therefore attribute the disordered areas of the surface to the presence of NCG.

We find that the amount of NCG that forms is dependent on the starting wafer that we use. We have studied five different wafers, with 5 – 15 samples studied from each wafer. Excellent sample-to-sample reproducibility is found for samples cut from a given wafer. Some wafers produce large amounts of NCG and some little; the former display in their morphology (after H-etching or graphitization) a relatively large density ($>1\times10^4$ cm$^{-2}$) of hexagonally-shaped etch pits as observed by optical microscopy, whereas the latter display very few such pits. This etch pit density is comparable to the density of threading dislocations expected in all of the wafers. We suspect that some difference in surface properties of the wafers produces the etch pits after H-etching on certain wafer but not on others. This variation in surface properties may also account for the differences in efficacy of graphene formation on the C-face found by other groups.[7,9]

In summary, we are able to make high quality graphene films on the C-face of SiC by annealing in vacuum. The morphology of these films differs from the ones grown on the Si-face, and we believe that this difference arises because of a differing interface structure between the substrate and the graphene in the two cases. We observe that the terraced structure found on the C-face graphene films has *thinner* graphene on the upper terraces, consistent with an interface-dependent graphitization rate. The material quality of our C-face graphene is high, with field-effect mobilities exceeding 4000 cm$^2$/Vs, as discussed elsewhere.[10] Finally, we find in some cases the presence of NCG on our graphene films. The amount of NCG is wafer-dependent, with wafers displaying few etch pits following H-etching and having straight, well-ordered step arrays on their surfaces producing the least amount of NCG.

Discussions with S. Nie, R. Berechman, and M. Skowronski are gratefully acknowledged. This work was supported by the National Science Foundation (grant DMR-0503748), and by the Defense Advanced Research Projects Agency through contracts administered by the Air Force Research Laboratory (contract FA8650-08-C-7823) and United States Air Force (FA8721-05-C002). Use of the Center for Nanoscale Materials at Argonne National Laboratory was supported by the U.S. Department of Energy, Office of Basic Energy Sciences, (contract DE-AC02-06CH11357). Opinions are those of the authors and are not necessarily endorsed by the U. S. Government.

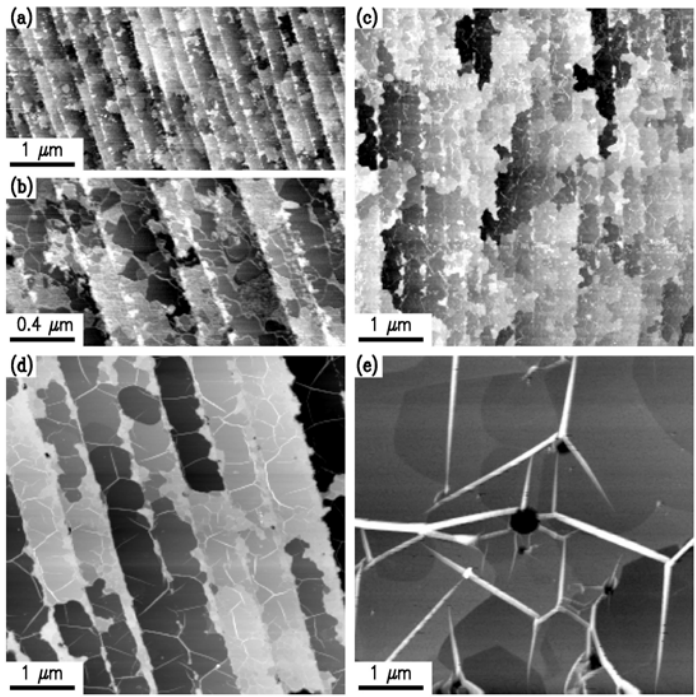

FIG. 1. AFM images of graphene formed on C-face 6H-SiC by annealing at temperatures of (and forming graphene thicknesses of): (a) and (b) 1120°C (1.2 ML), (c) 1190°C (4 ML), (d) 1240°C (9 ML), and (e) 1320°C (16 ML). Images are displayed with gray scale ranges of 2 nm, 2 nm, 4 nm, 13 nm, and 15 nm respectively.

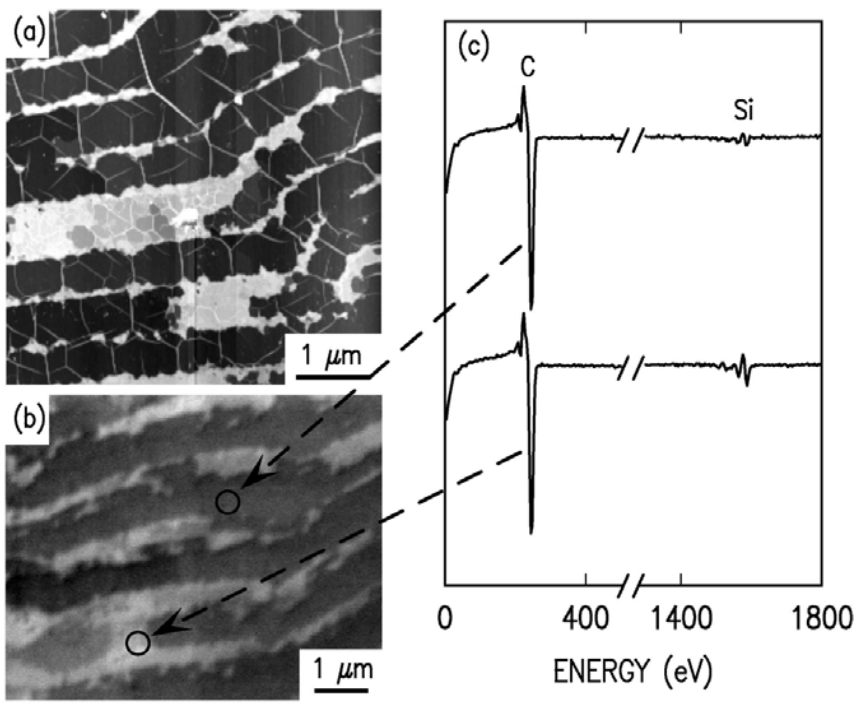

FIG.2. (a) AFM image of graphene on C-face 4H-SiC prepared by annealing at 1320°C, forming 8 ML of graphene. The image is displayed with a gray-scale range of 10 nm. (b) SEM image of same sample (different surface area). (c) Auger electron spectra acquired from the surface locations indicated in (b).

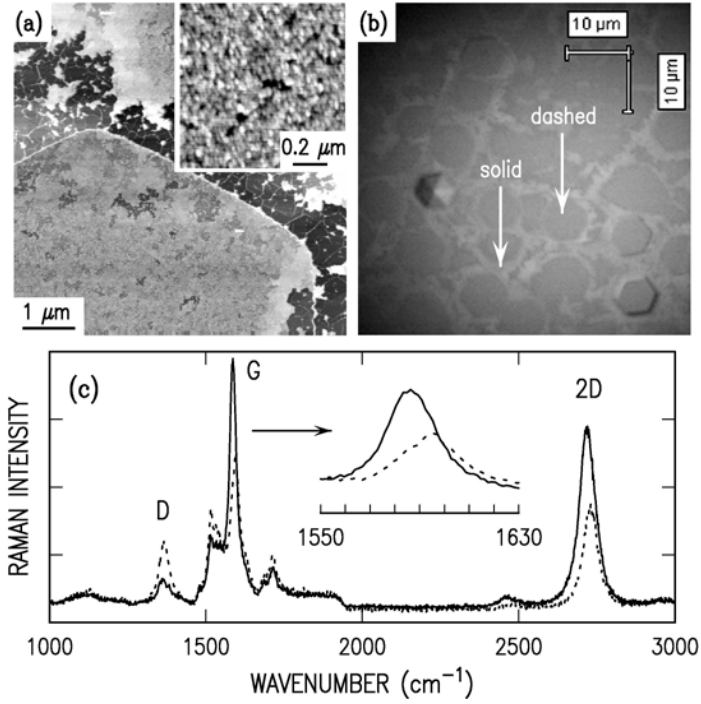

FIG. 3. (a) AFM image of graphene on C-face 6H-SiC prepared by annealing at 1220°C. The inset shows an expanded view of the higher (light gray) surface area. Gray-scale ranges are 9 nm for the main image and 2 nm for the inset. (b) Optical micrograph of same sample (different surface area). (c) Raman spectra, with the solid and dashed lines showing spectra acquired from the locations indicated in (b). The inset shows an expanded view of the G-line.

**References:**

[1] K. S. Novoselov, A. K. Geim, S. V. Morozov, D. Jiang, Y. Zhang, S. V. Dubonos, I. V. Grigorieva, and A. A. Firsov, Science **306**, 666 (2004).

[2] A. K. Geim, and K. S. Novoselov Nat. Mater. **6**, 183 (2007).

[3] J. M. Blakely, J. S. Kim, and H. C. Potter, J. Appl. Phys. **41**, 2693 (1970).

[4] A. J. Van Bommel, J. E. Crombeen and A. Van Toore, Surf. Sci. **48**, 463 (1975).

[5] L. Muehlhoff, W. J. Choyke, M. J. Bozack, and J. T. Yates, J. Appl. Phys. **60**, 2842 (1986).

[6] I. Forbeaux, J.-M Themlin, A. Charrier, F. Thibaudau, and J.-M Debever Appl. Surf. Sci **162/163** 406 (2000).

[7] J. Hass, R. Feng, T. Li, X. Li, Z. Zong, W. A. de Heer, P. N. First, E. H. Conrad, C. A. Jeffery, and C. Berger, Appl. Phys. Lett. **89**, 143106 (2006).

[8] K. V. Emtsev, F. Speck, Th. Seyller, L. Ley, and J. D. Riley, Phys. Rev. B **77**, 155303 (2008).

[9] N. Camara, G. Rius, J.-R. Huntzinger, A. Tiberj, L. Magaud, N. Mestres, P. Godignon, and J. Camassel, Appl. Phys. Lett. **93**, 263102 (2008).

[10] G. Gu, J. Kedzierski, Luxmi, N. Srivastava, P. J. Fisher, R. M. Feenstra, and Y. Sun, to be published.

[11] Luxmi, S. Nie, P. J. Fisher, R. M. Feenstra, G. Gu, and Y. Sun, J. Electron. Mater. **38**, 718 (2009).

[12] V. Ramachandran, M. F. Brady, A. R. Smith, R. M. Feenstra, and D. W. Greve, J. Electron. Mater. **27**, 308 (1997).

[13] J. Hass, F. Varchon, J. E. Millán-Otaya, M. Sprinkle, N. Sharma, W. A. de Herr, C. Berger, P. N. First, L. Magaud, and E. H. Conrad, Phys. Rev. Lett. **100**, 125504 (2008).

[14] F. Hiebel, P. Mallet, F. Varchon, L. Magaud, and J.-Y. Veuillen, Phys Rev B **78**, 153412 (2008).

[15] J. Hass, W. A. de Heer, and E. H. Conrad, J. Phys.: Condens. Matter **20**, 323202 (2008).

[16] C. Riedl, U. Starke, J. Bernhardt, M. Franke, and K. Heinz, Phys. Rev. B **76**, 245406 (2007).

[17] J. B. Hannon and R. M. Tromp, Phys. Rev. B **77**, 241404 (2008).

[18] A. C. Ferrari, J. C. Meyer, V. Scardaci, C. Casiraghi, M. Lazzeri, F. Mauri, S. Piscanec, D. Jiang, K. S. Novoselov, S. Roth, and A. K. Geim, Phys. Rev. Lett. **97**, 187401 (2006).

[19] A. Ferrari and J. Robertson, Phys. Rev. B **61**, 14095 (2000).